\documentstyle[aps,prd,preprint]{revtex}

\tightenlines

\newcommand{\be}{\begin{equation}}
\newcommand{\ee}{\end{equation}}
\newcommand{\bea}{\begin{eqnarray}}
\newcommand{\beas}{\begin{eqnarray*}}
\newcommand{\eea}{\end{eqnarray}}
\newcommand{\eeas}{\end{eqnarray*}} 
\newcommand{\ba}{\begin{array}}
\newcommand{\ea}{\end{array}}
\newcommand{\bi}{\begin{itemize}}
\newcommand{\ei}{\end{itemize}}
\newcommand{\ben}{\begin{enumerate}}
\newcommand{\een}{\end{enumerate}}

\newcommand{\la}{\langle}
\newcommand{\ra}{\rangle}

\newcommand{\seq}{\setcounter{equation}{0}}

\makeatletter
\def\vereq#1#2{\lower3pt\vbox{\baselineskip1.5pt \lineskip1.5pt
\ialign{$\m@th#1\hfill##\hfil$\crcr#2\crcr\sim\crcr}}}
\makeatother

\def\lsim{\mathrel{\rlap{\lower3pt\hbox{\hskip0pt$\sim$}}
    \raise1pt\hbox{$<$}}}         
\def\gsim{\mathrel{\rlap{\lower4pt\hbox{\hskip1pt$\sim$}}
    \raise1pt\hbox{$>$}}}         

\def\IC{\relax\hbox{$\inbar\kern-.3em{\rm C}$}}
\def\IQ{\relax\hbox{$\inbar\kern-.3em{\rm Q}$}}
\def\IR{\relax{\rm I\kern-.18em R}}

\def\IN{\relax{\rm I\kern-.18em N}}

 \font\cmss=cmss10 \font\cmsss=cmss10 at 7pt
\def\IZ{\relax\ifmmode\mathchoice
{\hbox{\cmss Z\kern-.4em Z}}{\hbox{\cmss Z\kern-.4em Z}}
 {\lower.9pt\hbox{\cmsss Z\kern-.4em Z}}
 {\lower1.2pt\hbox{\cmsss Z\kern-.4em Z}}\else{\cmss Z\kern-.4em Z}\fi}

\newcommand{\lL}{\varepsilon^{\ell_L}}

\newcommand{\hone}{\varepsilon^{H_1}}
\newcommand{\htwo}{\varepsilon^{H_2}} 

\begin{document}


\preprint{\vbox{ \hbox{UMD-PP-00-088} }}


\title{Theories in More than Four Dimensions\thanks{Lectures given at the
IX Mexican School on Particles and Fields, Metepec, Puebla, M\'exico,
August 9-19, 2000.}
}
\author{Abdel P\'erez-Lorenzana\thanks{email: aplorenz@glue.umd.edu}}
\address{
 Department of Physics, University of Maryland, College Park, Maryland 
20742, USA\\
 Departamento de F\'\i sica, 
Centro de Investigaci\'on y de Estudios Avanzados del I.P.N.\\
Apdo. Post. 14-740, 07000, M\'exico, D.F., M\'exico.}

\date{August, 2000}

\maketitle

\begin{abstract} 

Particle physics models where there are large hidden extra dimensions
are currently  on the focus of an intense
activity.  The main reason is that these large extra dimensions may come
with a TeV scale for quantum gravity (or string theory) which  leads to a
plethora of new observable phenomena in colliders 
as well in other areas of
particle physics. Those new dimensions could be as large as
millimeters implying deviations of the Newton's law of gravity at these
scales. 
Intending to provide a basic introduction to this fast developing
area, we present a general overview of theories with large extra
dimensions. We center our discussion on models for  neutrino masses, high
dimensional extensions of the Standard Model and  gauge coupling
unification. 
We discuss  the recently proposed technic of splitting fermion wave
functions
on a tick
brane which may solve the problem of a fast proton decay and produce fermion
mass hierarchies without invoking extra global symmetries. 
Randall-Sundrum model and some current trends  are also commented.

\end{abstract}


%
%
\section{Introduction}
\seq

New extra dimensions beyond the four of our world could possible exist in
Nature.  This idea is as old as  Kaluza and Klein's work dating back to
the  1920's~\cite{kk}. In modern terms, the idea  arose again with the
advent of string theories. However, it was conventional to assume that such
extra dimensions were compactified to manifolds  of small radii with a
size  of the order of the inverse Planck scale,   
$\ell_P=M_{P\ell}^{-1}=G_N^{1/2}$ or so, such that they would remain
hidden to low energy physics considerations.  Thus, beyond the interest of
a small community, the study of theories with extra dimensions  was
almost  far away from the scope of many particle physicists.  

During the last two  years of the XX century, the work on theories in more
than four dimensions  has increased almost exponentially.  The intriguing
fact that strongly  motivates this renewed interest is the realization of
the possibility that extra dimensions as large as
millimeters~\cite{dvali} could exist  and yet  being
hidden to the experiments~\cite{expt,feyn,coll1,sn87,coll2}, but with new
effects not so far of being observed.  Among the experimental signatures
are deviation on  the gravitational Newton's law  at small distances and a
rich phenomenology for collider physics. One  of the  attractive features
of these theories is that they  may provide a natural solution to the
hierarchy problem. In this scenarios, it is believed that we live on a
hypersurface  (3-brane) embedded in a higher dimensional world (the bulk). 
Although it is fair to say that  similar ideas  were  proposed on the 80's
by several authors~\cite{rubakov}, they were missed by some time, until
recent developments on string theory provided an independent realization
of such models~\cite{anto,witten,lykk}, given them certain credibility.

Our goal for the  present notes is  to give a general  overview of
this field.  Among the  scenarios presented until now in the literature,
we  will focus on studying mainly  the case of large (millimetric) extra
dimensions.  This, in turn, will provide us the insight to extend our
study to other more elaborated models. 

To motivate the ideas we  will start with the discussion of the origin of
the long standing hierarchy problem and how extra dimensions offer a new
way to understand it by providing a new  low fundamental scale, $M$,
which, based on the phenomenological and experimental constrains, could be
just at the TeV range.  At this point we shall assume that the
Standard Model particles are attached to the brane and that only gravity
propagates on the bulk.  As we will see, in such a  theory the weakness
of gravity is related to the large size of the volume of the extra space.

As it is clear, with a low fundamental scale all the particle physics
phenomena that invoke high energy scales will not work any more. Then,
problems as neutrino masses and mixings, gauge coupling unification and
proton decay should be reviewed under the light of this new theories. This
will be the central issue along our discussion, and in order to address
it we will modify  the above  model accordingly.

First, after discussing some general aspects of the brane-bulk theories,  
we will  explore neutrino oscillations phenomena. Here, we  shall
show how  an  isosinglet bulk  neutrino   that couples to
the  standard  neutrinos may  help to solve the neutrino
puzzles~\cite{nus,dienes1,dvali2,mnp,mp2,barbieri,pospel,3nu,valle2,beta,mp}. 
Next we  will analize the possibility that the standard model particles 
propagate in the extra dimensions and develop Kaluza-Klein
excitations~\cite{sm,quiros,earlier,mirabelli,ddg}.
The contribution of this
exited modes to standard processes will set  bounds to the bulk
radius and provide  collider signatures~\cite{coll2}.  
Those modes will also 
modify the profile of the renormalization group equations that
govern the running of the gauge coupling constants. The net effect is a
power law running  that  accelerates their  meeting, 
which now may occur at very low 
energies~\cite{ddg,kaku,uni,uni2,unimore,kubo,taylor}, even at the TeV
scale. 
We comment on 
the  accuracy of this low energy gauge coupling unification.

Next, we shall discuss a slightly modified scenario that may account for 
the explanation of fermion 
mass hierarchies and the suppression of proton decay by introducing a
splitting of the wave functions along the extra
dimension~\cite{sch1,sch2}.  Finally, we will present
the Randall-Sundrum
model~\cite{rs}
which provides an alternative explanation for the hierarchy problem based
on small extra dimensions and  a non factorizable bulk geometry. 
We will also  mention some 
of the current trends on the study of this class of 
models~\cite{gw,csaki,rsth,rizzo,grossman,cc,binetruy,cline,cosmo,cosmo2,perturb}.


\section{Hierarchy problem and large extra dimensions}
\seq
\subsection{Radiative corrections and the Higgs mass}

Our starting point is  the scalar sector  of the Standard Model
(SM). This is  perhaps the most peculiar 
sector in the theory. The scalar 
Higgs field, $H$, 
realizes the spontaneous symmetry breaking and gives masses to
all other particles, fermions and gauge bosons,  and yet, its own mass is
introduced as a free parameter on the theory. The Higgs is the only
particle of the SM which remains to be observed.  It is also the only
field that have self interactions that are only constrained by 
gauge invariance
and  the renormalizability condition which leave 
a $\lambda H^4$ term, with $\lambda$ a free parameter.  
And, what is more important for us, it is the
only field for which the quantum corrections require a large fine tunning
on the mass parameter. While the self energy diagrams of  all other
fields, evaluated in the $\overline {MS}$ scheme,  
develop logarithmic divergences
that depend on their own bare mass, the scalar field develops quadratic
divergences that are independent of its  bare mass. 
For instance, at one loop order one
gets 
 \be
 \delta m_H^2
 ={1\over 8\pi^2}\left(\lambda_H^2-\lambda_F^2\right) \Lambda^2
 + (\mbox{log. div.})
 + \mbox{finite terms.}
 \label{quads}
 \ee
where $\lambda_{H}$ is the self-couplings of $H$,
$\lambda_{F}$ is the Yukawa coupling  to fermions, and 
$\Lambda$ is the physical cut-off of the theory, which  is usually
believed to be the Planck scale, $M_{P\ell}\sim 10^{19}~GeV$, or the GUT
scale, $M_{GUT}\sim 10^{16}~GeV$.  


Nevertheless,  in order to keep the $WW$ scattering cross section from
violating unitarity, the physical Higgs boson mass, $m_H$, must be less
than about $1~TeV$. Then we get  the unpleasant result,
$m_H^2=m_{H,0}^2+\delta m_H^2+{\hbox{counterterm}}$, where the counterterm
must be adjusted to a precision of roughly $1$ part in $10^{15}$ in order
to cancel the  quadratically divergent contributions to $\delta m_H^2$.
Moreover, this adjustment must be made at each order in perturbation
theory. This large fine tunning is what  is known as the  hierarchy
problem.

Of course, the quadratic divergence can be renormalized away in exactly
the same manner  as is done for logarithmic divergences, and in principle,
there is nothing formally wrong with this fine tuning. In fact 
if this calculation is performed  in the
dimensional regularization scheme, $DR$, one obtains only $1/\epsilon$
singularities which are absorbed into the definitions of the counterterms,
as usual.  Hence, the problem of quadratic divergences does not become
apparent there.  It arises  only when one attempts to give a physical
significance to the cut-off $\Lambda$. In other words, if the SM were a
fundamental theory then the using of $DR$ would be justified. However, 
most theorists believe that the final theory should also include gravity,
then a cut-off must be introduced in the SM.
Hence, we regard this fine tunning as unattractive.

\subsection{A low fundamental scale from new dimensions}

Explaining the hierarchy problem  has been a leading motivations to
explore new physics during the last twenty years.
Supersymmetry~\cite{susy} and compositeness~\cite{tech} are  two of the
main proposals that solve this problem.   Supersymmetry predicts the
existence of new particles, the super-partners, at the TeV scale.  They
belong to the same dimensional representation than those of the SM, but
they differ each other on the spin. Their couplings are symmetric such that 
their
contribution to the quadratic divergences get balanced and  nullify
each other. This has been, so far,  the most popular extension of the SM. 
Compositeness, on the other hand, assumes that the Higgs is not a
fundamental field, but a quantum condensate of other fields.    
Both theories reflect
a common idea on the  structure of physics beyond the SM:  A new effective
field theory will be revealed at the weak scale, $m_{EW}\sim 1~TeV$,
stabilizing and perhaps explaining the origin of the hierarchy,
$m_{EW}/M_{P\ell}$, while, a desert among those scales will remain until
the Planck scale, which is assumed to be the fundamental scale where 
gravity becomes as strong as the gauge interactions and where the 
quantum theory of gravity is revealed. Eventually
the desert could be populated by other effective theories, responsible
from explaining other phenomena (as fermion masses) or 
from triggering dynamical
symmetry breakings, and a big deal of work has been dedicated to study
those pictures.

It has been realized recently that if there are more than four dimensions,
as it is suggested by string theory,  the above scenario could be
drastically modified.  To explain this let us assume as in
Ref.~\cite{dvali} that there are $\delta$  
extra space-like dimension which are
compact. Compactification will be in principle a natural ingredient
that would explain why we see only four dimensions.
Let us also imaging that all (some) of these dimensions are large with a
common radius $R$. Then, if  two test particles  with masses $m_1$ and 
$m_2$ were
separated each other  by a  distance $r \gg R$, they  would feel the usual
gravitational potential
 \be 
 U(r) = G_N{m_1 m_2 \over r}. \label{ur1}
 \ee
However, if $r \ll R$, the potential between each other should be
\be
U(r)= G {m_1 m_2\over r^{\delta+1}},
\ee
where $G^{-1}=M^{\delta+2}$, is the  coupling constant of gravity in $\delta+4$
dimensions which defines  the {\it fundamental scale} $M$ where gravity
becomes strong. Therefore, if we take the last relationship as the
fundamental one, then   Eq. (\ref{ur1}) will imply that
our {\it effective} four dimensional  gravity scale is given
by~\cite{dvali}
 \be
  M_{P\ell}^2 = M^{\delta+2} V_\delta  ,
  \label{mp}
 \ee
where $V_\delta$ is the volume of the extra space.  It is worth saying that
this same relationship arises if we  dimensional reduce to four
dimensions  the gravity action in $4+\delta$ dimensions, assuming   the
space-time to be ${\cal R}^4 \times {\cal M}_\delta$,  
where ${\cal M}_\delta$ is an
$\delta$ dimensional compact manifold of volume $V_\delta$. 
If the volume were large
enough, then the fundamental scale could be as low as $m_{EW}$, and the
hierarchy would be naturally removed. Of course, the price one has to pay
is to explain why the extra dimensions are so large. 
For $\delta=2$,  we get for
$M\sim 1~TeV$ that $R$ is less than 1 millimeter. This case
is highly interesting since it is on  the current limits on the low
distance gravity experiments~\cite{expt} 
that should detect the predicted deviation
of the  inverse squared law of gravity at distances, $r\sim
V_\delta^{1/\delta}$. Nevertheless, more than two extra dimensions  
could be expected
(strings predicts six more), and their size do not have to be
the same. More complex scenarios with a hierarchical distributions of the
sizes could be natural. Then, for the investigation of the model, we
will use a single large extra dimension,  implicitly
assuming other smaller dimensions.

Now well, while submillimeter dimensions remain untested for gravity, the
SM gauge forces have certainly been accurately measured up to weak scale
distances.  Therefore, the SM particles can not freely propagate in these
large extra dimensions,  but must be constrained to a four dimensional
submanifold. Then the  scenario  we have in mind is one where we live in a
four dimensional surface embedded in a higher dimensional space.  This
picture is similar to the D-brane models~\cite{dbranes},  as in the
Horava-Witten theory~\cite{witten}. We may also imagine our world as a 
domain wall of size $M^{-1}$ where the SM fields are trapped by some
dynamical mechanism~\cite{dvali}.  As this framework solves the
hierarchy problem,  supersymmetry is no longer needed. However, we should
notice that it may still be crucial for the self-consistency of the theory
of quantum gravity above the $M$ scale, although  such theory is yet
unknown. It could be  superstring theory, but it may also be something
yet to be discovered. In any case, supersymmetry  may coexist with 
large extra dimensions.

\subsection{Experimental bounds}

There are a number of dramatic experimental consequences of large extra
dimensions. First, as already mentioned, there is the deviation of the
inverse squared law on gravity at submillimeter distances. The current
experiments are just at this limit, so $R<0.2~mm$~\cite{expt}. 
Also, as gravity
becomes comparable in strength to the gauge interactions at energies $M
\sim$ TeV, the nature of the  quantum theory of gravity would become
accessible to  LHC and NLC.  With gravity freely propagating on the bulk, 
the effect of the gravitational couplings will be mostly of
two types: missing energy, that goes into the bulk, and corrections to the
standard cross sections from graviton exchange.  The Feynman rules were
calculated from the linearized bulk gravity theory, and  are
given   in Ref.~\cite{feyn}. From the effective four dimensional point
of view,  the graviton  develops Kaluza Klein (KK) excitations of
masses $n/R$, with $n$ an integer number (see next section). The 
coupling of each one of those modes  to the SM particles is  
suppressed by the Planck scale, but the overall coupling that considers
all the KK excitations become suppressed  just by $M$.  

A long number of studies on this topic have appeared
already~\cite{feyn,coll1},  some nice and short reviews of collider signatures
are given in~\cite{review1}. We summarize some of the current bounds  in
tables 1 and 2. At $e^+e^-$ colliders (LEP,LEPII, L3), the best signals
would be the production of gravitons with $Z,\gamma$ or fermion pairs
$\bar ff$.  There is also the interesting  monojet
production~\cite{feyn} at hadron colliders (CDF, LHC) which is yet
untested. 

The virtual exchange of gravitons  either leads to modifications of
the SM cross sections and asymmetries, or to new processes not allowed in
the SM at tree level. The amplitude for exchange of the entire tower
naively diverges when $\delta>1$ and has to be regularized, typically by 
a cut
off at $M$, which performs the replacement of the sum suppressed by
$M_{P\ell}$ into the simple suppression by $M$. An interesting channel is
$\gamma\gamma$ scattering, which appears at tree level, and may surpasses
the SM background at $s=0.5$ TeV for $M= 4$ TeV. Bi-boson productions of
$\gamma\gamma$, $WW$ and $ZZ$ has been already analized~\cite{feyn,coll1}.
Some
experimental limits, most of them based on existing data, are given
in Table 2. The upcoming experiments will 
easily overpass those limits. 

Notice that all collider limits are about 1 TeV. A more stringent
constraint comes from SN1987A and astrophysics~\cite{sn87}, which gives
$M>30 - 100$ TeV.

%
%

\section{An introduction to brane-bulk models}
\seq

Before getting into the  discussion of models with large extra dimensions
in more detail, let us make some useful remarks on the generical
properties of brane-bulk models.

As we already mentioned, from the four dimensional effective theory point
of view, a bulk field, as the graviton,  will appear as an infinite tower
of KK excitations. The appreciation of the impact of this KK excitations
will depend in the relevant energy of the experiment, and on the
compactification scale $1\over R$. Roughly speaking, if $E\ll {1\over R}$
the theory behaves purely four dimensional. However, for energies above 
$1\over R$, a large number 
of  KK excitations, $\sim (ER)^\delta$, becomes kinematically
accessible making physics looks $4+\delta$ dimensional. 
To clarify this matters let us consider a five dimensional model where the
fifth dimension has been compactified on a circle of circumference $2\pi
R$. The generalization of these results is straightforward.
Let $\phi$ be  a  bulk scalar field  for which the action
has the form 
 \be  
 S_\phi = {1\over 2}\int\! d^4x\ dy\  
 \left(\partial^A\phi\partial_A\phi  - m^2 \phi^2\right) ;
 \ee
where $A=1,\dots,5$, and $y$ denotes the fifth dimension. 
Demanding periodicity on the extra compact dimension under 
$y\rightarrow y+2\pi R$,
the field may be Fourier expanded as 
 \be 
 \phi(x,y) = {1\over\sqrt{2\pi R}}\phi_0(x) + 
 \sum_{n=1}^\infty {1\over\sqrt{\pi R}} 
 \left[ \phi_n(x)\cos\left({ny\over R}\right)
  + \hat\phi_n(x)\sin\left({ny\over R}\right)\right].
 \label{kkexp}
 \ee
 Notice  the different normalization of the excited modes, $\phi_n$ 
and $\hat\phi_n$, with respect to the zero mode, $\phi_0$.  By introducing
the last equation into the action and integrating over the extra
dimension we get
 \be 
 S_\phi = \sum_{n=0}^\infty {1\over 2}\int\! d^4x 
 \left(\partial^\mu\phi_n\partial_\mu \phi_n - m_n^2 \phi_n^2\right)
 +  \sum_{n=1}^\infty {1\over 2}\int\! d^4x 
 \left(\partial^\mu\hat\phi_n\partial_\mu\hat\phi_n - 
 m_n^2\hat\phi_n^2\right),
 \ee
where the KK mass is given as $m_n^2 = m^2 + {n^2\over R^2}$. Therefore,
in the effective theory, the higher dimensional field looks like an
infinite tower of fields with masses $m_n$. These are the so called  KK
modes.  Notice that all these modes are fields with the same spin, and
quantum numbers as $\phi$.  But they differ in the KK number $n$,
associated with the fifth component of the momentum. For $m=0$
it is clear that
for energies below $1\over R$ only the massless zero mode
will be kinematically accessible, making the theory looks four dimensional.
As the energy increases, once it 
surpasses the threshold of the first exited
level, the manifestation of the KK modes will evidence the higher
dimensional nature of the theory.

The five dimensional field $\phi$ has mass 
dimension $3\over 2$, while all
modes are dimension one. In general for $\delta$ extra dimensions we will get 
$[\phi] = d_4 + {\delta\over 2}$, where $d_4$ is the natural mass
dimension of  the
field in four dimensions. Because this change on the dimensionality of
$\phi$, the higher order operators (beyond the mass term) 
will all have dimensionful couplings. 
To keep them dimensionless a mass parameter should be  introduced to
correct the dimensions. As usual, the natural choice for this parameter is
the cut-off of the theory. For instance, let us consider the quartic
couplings of $\phi$. Since all potential terms should be of dimension
five, we should write down ${\lambda\over M} \phi^4$.  After
integrating the fifth dimension, this operator will generate quartic 
couplings among all KK modes. Four normalization factors containing
$1/\sqrt{R}$ appear in the expansion of $\phi^4$. Two of them will be 
removed by the integration, thus,  we left with the  effective
coupling $\lambda/MR$.  By introducing Eq. (\ref{mp}) we  observe that the
effective couplings have the form
 \be 
 \lambda\left( M\over M_{P\ell}\right)^2 \phi_k\phi_l\phi_m\phi_{k+l+m}.
 \ee
where the indices are arranged to respect the conservation of the fifth
momentum.  From the last expression we conclude that in the low energy
theory ($E<M$),  the effective coupling appears suppressed. Thus, the 
effective four dimensional theory is weaker interacting compared with the
bulk theory.  Something similar happens to gravity on the bulk, where  the
coupling constant is stronger than in the brane, because it is also
suppressed by the volume of the extra space as given in Eq. (\ref{mp}).  By
the naturalness principle, we  must assume that all dimensionless 
coupling constants are of order one.

Let us now consider the  brane fields
represented by a  fermion  $\psi(x)$.  
The theory describing this brane fermion  is purely four
dimensional. Indeed, its action  is 
localized by a delta function in the
complete theory in the form 
 \be 
 S_{\psi} = \int\! d^4x\ dy\ {\cal L}\left(\psi\right)\delta(y-y_0),
 \ee
where $y_0$ is the position of the brane.
The coupling of $\psi$ to the bulk scalar field only 
may take place at the brane, where $\psi$ lives. For simplicity we
will assume that the brane is located at the position $y_0=0$, 
which in the case of orbifolds corresponds  to a fixed point. 
So, the part  of the action that describes the brane-bulk coupling 
is
 \be 
 S_{int} = \int\! d^4x\ dy\ {\cal L}_{int}\left(\phi,\psi\right)\delta(y).
 \ee
Lets choose for instance  the term
 \bea
 S_{\phi\psi} &=& 
 \int\! d^4x\ dy\ {f\over \sqrt{M}}\bar\psi(x)\psi(x)\phi(x,y=0)\ \delta(y)
  \nonumber\\
 &=& \int\! d^4x {M\over M_{P\ell}}f \cdot \bar\psi\psi
 \left(\phi_0 +  \sqrt{2}\sum_{n=1}^\infty  \phi_n\right).
 \label{Lint}
 \eea
Here the Yukawa coupling constant $f$ is dimensionless. On the right hand
side we have used the expansion (\ref{kkexp}) and Eq. (\ref{mp}). From
here, we notice that the coupling of brane to bulk fields is generically
suppressed by the ratio $M\over M_{P\ell}$. Also, notice that   the modes
$\hat\phi_n$  decouple from the brane. Then,  to get ride of those
(harmless) fields in the effective theory we could assume that $\phi$ is
an even field under the transformation $y\rightarrow -y$, thus, imposing a
${\cal Z}_2$ symmetry on the theory, though the theory is consistent
even without this extra assumption. That  symmetry, on the other hand, is
characteristic of the orbifolds and domain walls. By picking up an
explicit parity for $\phi$ we are projecting out one half of the KK
excitations, it means, the expansion (\ref{kkexp}) will involve either the
cosine or the sine part, but not both. 

Let us stress that the couplings in (\ref{Lint}) do not conserve  the KK
number. 
This reflects the fact that the   brane  breaks
the translational symmetry along the extra dimension. Nevertheless, it is
worth noticing that the four dimensional  theory is still Lorentz
invariant. Physically this  means that when the interactions among the
brane fields reach enough energy to produce real emission of  KK
modes, part of the energy of the brane is released into the bulk.
This is also the case of gravity.
 
Let us mention, parenthetically, 
that in general the  linear perturbations of the metric
lead to particles of  spin two, one and zero; however, only the spin two
graviton, $G_{\mu\nu}$, and one scalar field, $b$, 
the radion; couple to matter at the weak
field limit. The Feynman rules are given in~\cite{feyn}. Briefly speaking,
they come from the  couplings to the energy momentum tensor
 \be 
 {\cal L} = -{1\over M_{P\ell}}\sum_n \left[G^{(n)\mu\nu} - 
 {\kappa\over 3}b^{(n)}\eta^{\mu\nu}\right] T_{\mu\nu}.
 \ee
Here $\kappa$ is a parameter of order one. Notice that $G^{(0)\mu\nu}$ is
massless. That is the source of long range four dimensional gravity. 
It is worth saying that $b^{(0)}$ is not actually massless, it get a mass
from the stabilization mechanism. From supernova constrains such a mass
should be larger than $10^{-3}$ eV.  Experimental
bounds for graviton production were given in the previous section.

Next, let us consider the scattering process among brane fermions:
$\psi\psi\rightarrow\psi\psi$. In the present toy model this interaction
is mediated by all the KK excitations of $\phi$, then the typical
amplitude will receive the contribution
 \be
 {\cal M} = \hat f^2  
\left( {1\over q^2 - m^2} +  2\sum_{n=1}{1\over q^2 - m_n^2}\right) D(q^2) ,
 \ee 
where $\hat f$ represents the  effective coupling, and $D(q^2)$ is an
operator that depends on the Feynman rules and is usually 
independent of the index $n$, as a consequence of the universal coupling
manifested on Eq. (\ref{Lint}).
In more than five dimensions the equivalent to the above sum usually
diverges and has  to be regularized by introducing a cut-off at the
fundamental scale.  Roughly speaking, at high energies, $qR\gg 1$, the
overall factor becomes $\hat f^2 N$, where $N$ is the number of KK modes
until the cut-off. It is $N=MR= M_{P\ell}^2/M^2$. In the case of the
graviton, $\hat f$ is about $1/M_{P\ell}$, and therefore, the overall
coupling becomes just $1/M$~\cite{feyn,coll1}. At low energies, on the
other hand, by
assuming that $q^2\ll m^2\ll 1/R^2$
we may integrate out the KK excitations, and at the
first order we get
 \be
 {\cal M}={\hat f^2\over m^2} D(q^2)\left(1+{\pi^2\over 3}m^2R^2\right).
 \label{msum}
 \ee
The last term between parenthesis is a typical correction produced by the
KK modes exchange to the pure four dimensional result.

Let us now mention on some characteristics of other bulk fields. 
A massless bulk fermion field is defined as the solution of the higher
dimensional Dirac Equation  $\left( i\partial^M\Gamma_M \right)\Psi=0$,
where the $\delta+4$ Dirac matrices satisfy the  algebra
$\{\Gamma_M,\Gamma_N \} = 2 g_{MN}$. In five dimensions we use the Weyl
basis
 \be
 \Gamma_\mu = \gamma_\mu = 
 \left(\ba{cc} 0&\sigma_\mu\\ \bar\sigma_\mu & 0\ea\right); 
 \qquad \mbox{ and } \qquad
 \Gamma_5 = \gamma_5 = 
  -i \left(\ba{cc} {\bf 1}&0\\ 0 & {\bf -1}\ea\right).
 \ee
Therefore, a bulk fermion is necessarily a  four component spinor. 
Thus, only
vectorlike theories are in principle possible.
However, if the fifth dimension is orbifolded, the theory on the brane may
look as a chiral theory.   In this basis, $\Psi$ is
conveniently  decomposed as
 \be
  \Psi = \left(\ba{c} \nu_R \\ \nu_L \ea \right);
 \ee 
with each component having its own set of KK excitations. 
The free action for the massless field may 
be expanded in terms of the KK modes, and  it has the form
 \bea 
 S_{\Psi} &=& \int\! d^4x \ dy\ i \bar\Psi\Gamma^M\partial_M\Psi
 \nonumber \\
  &=& \int\! d^4x\ \left(
 i\bar\nu_{nL}\bar\sigma^\mu\partial_\mu \nu_{nL} + 
 i\bar\nu_{nR}\bar\sigma^\mu\partial_\mu \nu_{nR}  - 
 {n\over R}\bar\nu_{nR} \nu_{nL} + h.c \right) ;
 \label{KKnu}
 \eea
where a sum over $n$ is to be understood. Notice that all KK masses are
Dirac masses. 

Two different Lorentz invariant fermion bilinears are possible in five
dimensions: Dirac mass terms $\bar\Psi\Psi$ and Majorana masses
$\Psi^TC_5\Psi$, where $C_5= \gamma^0\gamma^2\gamma^5$.  However, the
Dirac mass is an odd function under the orbifold symmetry, $y\rightarrow
-y$, under which $\Psi\rightarrow \gamma^5\Psi$. So, if the theory is
invariant under this symmetry that term will be zero. 

Lets now consider the extension of a gauge field to five dimensions. For
simplicity let us  consider only  the case of a free gauge abelian theory.
The Lagrangian in five dimensions is given as
 \be 
  {\cal L}_{5D} = -{1\over 4} F_{MN}F^{MN} \ .
 \ee
Upon integration over the extra dimension one gets~\cite{ddg}
 \be
 {\cal L} =-{1\over 4}\sum_{n=0}^\infty F_{\mu\nu}^{(n)}F^{(n)\mu\nu}
  - {1\over 2}\sum_{n=0}^\infty 
 \left[\partial_\mu A_5^{(n)} + {n\over R}A_\mu^{(n)} \right]^2
 \ee
The gauge invariance of the theory can be now expressed in terms of the
(expanded) gauge transformation of the KK modes
\be
  A_{\mu}^{(n)} \rightarrow 
  A_{\mu}^{(n)} + \partial_{\mu} \theta^{(n)} \qquad \mbox{ and } \qquad
  A_5^{(n)} \rightarrow A_5^{(n)} - {n \over R} \theta^{(n)} .
\ee
Therefore, by fixing the gauge 
one may absorb the scalar field $A_5$ into $A_\mu$. 
Hence, the only massless vector is the zero mode, all KK modes  acquired
mass by absorbing the Goldston bosons, $A_5^{(n)}$, 
associated to the  spontaneous
isometry breaking~\cite{dobado}. That keeps the gauge symmetry on the
four dimensional effective theory untouched. We must stress that this new theory is
essentially non renormalizable for the infinite number of fields that it
involves. However, the truncated theory that only consider a finite number
of KK modes is renormalizable. The cut-off for the inclusion of the
excited modes will be again the scale $M$. Other aspects of this theories
will be discussed later on when we addressed the SM extensions.

\section{Neutrino masses}
\seq

Several experiments have provided conclusive evidence for a deficit on the
expected flux of  atmospheric and solar neutrinos, and there is also  the
direct observation of $\bar\nu_e$ in the $\bar\nu_\mu $ beam of the LSND
experiment.  A simple explanation of these anomalies  arises if neutrinos
are massive and a large amount of work is devoted nowadays to explain the
origin of their small squared mass differences~\cite{nus}. However, most of
the four dimensional mechanisms invoke high energy physics with scales
about $10^{12}$ GeV or higher. Obviously, with a fundamental TeV scale, 
understanding the small neutrino masses poses a  theoretical challenge
to the new  theories. 

A second possible problem is the enhancement of all non renormalizable
operators, now suppressed only by powers of $1/M$.  Among them, there 
are  those  which produce 
a dangerous fast proton decay and 
the operator 
 \be  {(LH)^2 \over M} \ee
which produces a large neutrino mass of the order $\la H \ra^2/M$.  To
exclude this
operator  one has to make the additional assumption that the theory
respect lepton number symmetry (or more precisely $B-L$).  Two class of
models are then possible depending on whether $B-L$ is a global or local
symmetry. We discuss both possibilities in this section.

\subsection{Models with global $B-L$ symmetry}

In the context of models that have  a global $U(1)_{B-L}$ symmetry, one
can  get small neutrino masses by introducing isosinglet neutrinos in the
bulk~\cite{dienes1} which carry lepton number.  As this is a sterile
neutrino,  it comes natural to assume that it may propagate into the bulk
as well as gravity, while the SM particles remain attached to the brane.
These models are interesting since they lead to small neutrino masses
without any extra assumptions. 

Let  $\nu_B( x^{\mu}, y)$ be a bulk neutrino which we take to be massless 
since the Majorana mass violates conservation of Lepton number  and the
five dimensional Dirac mass is  forbidden by the orbifold symmetry,
which we assume. This neutrino couples  to the
standard lepton doublet, $L$, and  to the  Higgs field, $H$, via 
${h\over \sqrt{M}}\bar{L} H \nu_{BR}~\delta(y)$. Once the Higgs develops its
vacuum, this coupling will generate 
the four dimensional Dirac mass terms
 \be
  m\bar\nu_L\left( \nu_{0R} + \sqrt{2}\sum_{n=1}^{\infty} \nu_{nR}\right) ,
 \label{l1}
 \ee  
where the mass $m$ is given by~\cite{dvali2}
 \be 
 m = h v {M\over M_{P\ell}}\sim 10^{-4}~eV\times {hM\over 1~TeV}.
 \ee
Therefore, if $M\sim 1~ TeV$ we get just the right order of magnitude on the
mass as required by the experiments. Moreover,   even if 
the KK decouple for a small  $R$, we will still  get the
same Dirac mass for $\nu_L$ and $\nu_{0R}$, as far as $M$ remains in the
$TeV$ range.
After including the KK masses from Eq. (\ref{KKnu}), we may write down all
mass terms in the compact form~\cite{mnp}
 \be 
   (\bar{\nu}_{eL}  \bar{\nu}'_{BL})\left(\begin{array}{cc}
 m &\sqrt{2} m\\ 0 & \partial_5
 \end{array}\right)\left(\begin{array}{c}\nu_{0B} \\
 \nu'_{BR}\end{array}\right),
 \label{m1}
 \ee
where the notation is as follows:  $\nu'_B$ represents the KK
excitations.  The off diagonal term  $\sqrt{2} m$  is actually an infinite
row vector of the form $\sqrt{2} m (1,1,\cdots)$ and the operator
$\partial_5$ stands for the diagonal and infinite KK mass matrix whose
$n$-th entrance is  given by $n/R$.  
Using this short hand notation  it is straightforward  to
calculate the exact eigensystem for this mass matrix~\cite{mp2}. 
Simple algebra yields the characteristic equation
 $2 \lambda_n = \pi \xi^2 \cot(\pi \lambda_n)$,
with $\lambda_n=m_nR$, $\xi=\sqrt{2}mR$, and where $m_n$ is the mass 
eigenvalue~\cite{dienes1,dvali2}. 
The weak eigenstate is given in terms of the mass eigenstates, 
$\tilde \nu_{nL}$, as 
 \be
 \nu_L = \sum_{n=0}^\infty {1\over N_n} \tilde \nu_{nL} ,
 \label{nul}
 \ee
where the mixing $N_n$ is  given by
$N^2_n = \left(\lambda_n^2 + f(\xi)\right)/\xi^2$,
with $f(\xi)= \xi^2/2 + \pi^2 \xi^4/4$~\cite{mp2}.
Therefore, $\nu_L$ is actually  a coherent superposition of an
infinite number of massive modes. As they evolve
differently on time, the above equation will give rise to neutrino
oscillations, $\nu\rightarrow \nu_B$, 
even though there is only one single flavour. This is a
totally new effect that arise the possibility of new ways of understanding
the neutrino anomalies.  An analysis of the implications of the mixing
profile in these models  for solar neutrino deficit was presented
in~\cite{dvali2}.  Implications for atmospheric neutrinos were
discussed in~\cite{barbieri}, and some phenomenological bounds were given
in~\cite{barbieri,pospel}. A comprehensive analysis for 
three flavours is given in \cite{3nu}. Here we
summarize some of the main results.

\subsection{New patterns of neutrino oscillations}

{\it Small $\xi$}.
For $\xi\ll 1$, the mixing in Eq. (\ref{nul}) is given by
$\tan\theta_n\sim\xi/n$~\cite{dvali2}. The masses are: $m$ for $n=0$ 
and $n/R$ otherwise.
Therefore, as expected, the main component of $\nu_L$ is the lightest
mass eigenstate.  
The mixing, on the other hand, will induce a resonant conversion into bulk
(sterile) modes. The survival probability has the form
 \be
 P_{surv}(L) 
  = 1- {4\over \eta^4}\xi^2 \sum_{n=1}^\infty 
  {\sin^2\left({n^2 L\over 4ER^2}\right)\over n^2} - 
  {2\over \eta^4}\xi^4 \sum_{k,n=1}^\infty 
  {\sin^2\left[{(n^2-k^2)L\over 4ER^2}\right]\over n^2 k^2}, 
  \label{ps1}
 \ee
where $\eta= (1 + \pi^2 \xi^2 /6)^{1/2}$.
Thus,  the oscillation length is given by $L_{osc}= 4\pi ER^2$. 
Clearly, at the low $\xi$ order the above expresion becomes 
$P_{surv}(L) \approx 1- 4\xi^2 \sin^2(L/4ER^2)$, which mimics the
former small mixing angle case.
Therefore, if $1/R\sim 10^{-3}~eV$, which is just $R\sim 0.2~mm$,
we get an explanation for  solar
data by introducing MSW effects as described in~\cite{dvali2} that is
consistent with other 
astrophysical constrains~\cite{dvali2}, though  it has been suggested that 
SN1987A may impose
more stringent bounds: $R<1$ \AA~\cite{barbieri}.
A similar explanation for the atmospheric anomaly seems
disfavored~\cite{dvali2,3nu}, since it needs large mixing angle.

{\it Large $\xi$}.
On the continuos approximation, 
the survival probability reads~\cite{barbieri}
  \be
 P_{surv}(z)  =
 \left|1 - {\rm erf}\left({\pi\over 2}\xi^2\sqrt{iz}\right)\right|^2  
 \label{ps2}
 \ee
where $z= L/2ER^2$. 
Notice that now the probability has no oscillatory nature in terms of
$L/E$, this feature should distinguish this models form the conventional
two neutrino oscillations. Indeed, it is
easy to check the physical origin of this effect 
from the exact solutions~\cite{3nu}. For
$\xi\gsim 1$, the eigenvalues $\lambda_n$ start to deviate from the integral
value $n$, and a large number ($N_\xi\sim \pi^2\xi^2/4$) 
of equally  suppressed eigenstates
contribute to (\ref{nul}).
Then, once $\nu_L$ is released, the 
time evolution of the different
components will most likely wash out the original coherent superposition 
and the initial $\nu_L$ will almost disappear, and 
the  maximal probability will not be recovered away from the
source~\cite{dienes1}. 
The fast developing slope of $P_{surv}$ is governed by 
the single parameter $\xi^2/R$. As solar and atmospheric have
$\overline P_{surv} \approx 0.5$, in order 
to avoid a large deficit on $P_{surv}$, one must keep 
that parameter on a small range. 
For atmospheric one gets that 
$\xi^2/R\approx 10^{-2} eV$~\cite{barbieri,3nu}. That means 
$10^{-3}~eV<1/R<10^{-2}~eV$, and $1<\xi^2<10$, and then, 
$R$ should remain in the submillimeter range. An explanation for solar
data is,  on the other hand, not possible in this limit~\cite{3nu}.

\subsection{Three flavour oscillations} 

The extension of this model to three
brane  generations, $\nu_{e,\mu,\tau}$, is straightforward.  
It was observed
earlier~\cite{mp2} that to give masses to the three standard generations
three bulk neutrinos are needed. This comes out 
from the fact that with a
single bulk neutrino only one massless right handed neutrino is present
(the zero mode), then, the coupling to brane fields will generate only one
new massive Dirac neutrino.  After introducing a rotation by 
an unitary matrix
$U$ on the weak sector, the most general Dirac  mass terms with three
flavours and arbitrary Yukawa couplings may be written down as 
 \be 
 -{\cal L} =
 \sum_{\alpha=1}^3\left[ 
  m_\alpha \bar\nu_{\alpha L} \nu^\alpha_{BR}(y=0) + 
  \int dy\, \bar\nu^\alpha_{BL}\partial_5 \nu^\alpha_{BR} + 
  h.c. \right] ,
 \label{three} 
 \ee
where
$\nu_{aL}=U_{a\alpha}\nu_{\alpha L}$, 
with $a= e,\mu,\tau$ and $\alpha=1,2,3$. 
The  mass parameters $m_\alpha$ are the
eigenvalues of the 
Yukawa couplings matrix multiplied by the vacuum $v$, and as stated before
are naturally of the order of eV or less.
This reduces the analysis to
considering three sets of mixings given as in the previous case. Each set
(tower) of mass eigenstates is characterized by
its own parameter $\xi_{\alpha}\equiv \sqrt{2}m_{\alpha} R$. Now, each
weak eigenstate can be expressed as a coherent superposition of these
three different towers by
 \be
 \nu_a = 
  \sum_{\alpha=1}^3 \sum_{k=0}^\infty 
  U_{a\alpha} {1\over N_{\alpha k}} \tilde\nu_{\alpha k} ;
 \label{gennu}
 \ee
with $\tilde\nu_{\alpha k}$ being the $k$-th 
mass eigenstates of the $\alpha$-th
tower.
It is now clear that  the
three flavour oscillations will  correspond to the oscillations among the
three towers. In this regards, if the KK do not decouple, 
the explanation to neutrino puzzles will not be 
any more described  in terms of three single neutrinos. 
Now, the transition probability is given by
 $P_{ab} =  \sum_{\alpha\beta} 
 U^*_{a \alpha}U_{b\alpha}U^*_{b\beta}U_{a\beta}\  p_{\alpha\beta}$,
where we have introduced the partial transition 
probabilities $p_{\alpha\beta}$ defined as
 $p_{\alpha\beta}(L) \equiv
 \overline {\langle\nu_\alpha(L)|\nu_\alpha(0)\rangle }
 \langle\nu_\beta(L)|\nu_\beta(0)\rangle$. The diagonal 
 $p_{\alpha\alpha}$ is interpreted as the survival 
probability of (the non standard) $\nu_\alpha$, and it has the form of
(\ref{ps1}) and (\ref{ps2}) for small and large $\xi_\alpha$ respectively.

The analysis of Ref.~\cite{3nu}  shows that there is not simultaneous
explanation for solar, atmospheric and LSND data within  this minimal
model. Without LSND, we  have  the following possible
scenarios: 
(i) $\xi_{1,2,3}\ll 1$; then, $R$ is
smaller than $1~\mu m$. KK modes decouple. 
Solar and atmospheric
neutrino data are understood as in the case of four dimensional models.
(ii) $\xi_{1,2}\ll 1\lsim \xi_3$; solar neutrino data is provided
as in the four dimensional models but atmospheric data is explained by
$\nu_{\mu}$ to $\nu_{bulk}$ oscillation as from Eq. (\ref{ps2}).
 $R$ is on the submillimeter range.
(iii) $\xi_1\ll 1 \lsim \xi_{2,3}$. Both, solar and atmospheric
data are explained by  $\nu\rightarrow\nu_{bulk}$ oscillations. 
$1/R^2\sim \Delta m_{sol}^2\sim 10^{-3}~eV$ 
with matter effects for solar and  $\xi_{2,3}^2\sim 10$. Finally,
(iv) for $\xi_{1,2,3}\gg 1$ there is no explanation for solar
neutrino data. This case is therefore ruled out.  

\subsection{Models for Majorana masses}

Some extended scenarios that consider the  generation of  Majorana masses
from the breaking of lepton number either on the bulk or on 
a distant brane have been
considered in Refs.~\cite{valle2,beta} 
(the breaking of  global symmetries at distant branes 
was first proposed in Ref.~\cite{break}). In these models 
a bulk scalar field  $\chi$ that carries lepton number (2)
is introduced. It develops a small vacuum and gives mass to the neutrinos
which are generically of the form
\be 
m_\nu \sim {\la H\ra^2\over M} {\la\chi\ra_B\over M^{3/2}};
\ee
then, with 
$M$ of the order of 10 TeV, we need $\la\chi\ra_B\sim (10~ MeV)^{3/2}$. 
Such a small vacuum is possible in both classes of models, 
though it usually needs a small mass for $\chi$. 
Obviously, with Majorana masses, a bulk
neutrino may not be needed but new physics must be invoked. 
We should notice the there is also  a Majoron field associated to the
spontaneous breaking of the lepton number symmetry.
Its phenomenology depends on the details of the specific model.  
In the simplest scenario, the coupling $(LH)^2\chi$ 
is the one responsible for generating Majorana masses. 
It also gives  an important contribution for neutrinoless
double beta decay which is just right at the current  experimental 
limits~\cite{beta}.

\subsection{Models with local $B-L$ symmetry}

We now proceed to consider the second class of models. For the case where
$B-L$ is local,  anomaly cancellation requires that right handed
neutrinos must be present in the brane as in the models discussed in
Ref.\cite{mnp,mp2}. 
The simplest gauge
model where this scenario is realized  is the left-right symmetric model
where the right handed symmetry is broken by the Higgs  doublet 
$\chi_R(1,2,1)$, where the number inside the parenthesis correspond to the
quantum numbers under $SU(2)_L\times SU(2)_R \times U(1)_{B-L}$.
The model then contains the left and right handed brane leptons and a blind
bulk neutrino.
The relevant terms of the  action for one generation are~\cite{mnp} 
 \[
 {\cal S}= \int d^4x [\kappa\bar{L}\chi_L \nu_B(y=0) +
 \kappa \bar{R}\chi_R
 \nu_B(y=0)+h \bar{L}\phi R] + \int d^4x dy
 \bar{\nu}_B\Gamma^5\partial_5\nu_B + h.c.
 \]
By setting  $\la\chi^0_R\ra= v_R$ and
$\la\chi^0_L\ra=0$, the following Dirac neutrino mixing  matrix is obtained
 \be
 (\bar{\nu}_{eL}~\bar{\nu}_{0BL}~  \bar{\nu}'_{BL})\left(\begin{array}{cc}
 hv & 0\\ \kappa v_R & 0 \\\sqrt{2}\kappa v_R &  \partial_5
 \end{array}\right)\left(\begin{array}{c}\nu_{eR} \\
 \nu'_{BR}\end{array}\right).
  \label{mright}
 \ee
Note that in general the effect of 
$\la\chi^0_L\ra=0$ may also be produced if 
the bulk neutrino breaks explicitly the parity symmetry~\cite{mp} (as it
does if orbifold symmetry is present).
Now, 
a massless field, $\nu_0$,  which is predominantly the electron
neutrino, provided that $\kappa v_R\gg hv\simeq$ few MeV, appears. Since 
$\kappa\simeq \frac{M}{M_{P\ell}}$, this constraint implies that $M$ 
must be as large as $10^8$ GeV or so, however $R$ may remain in the
submillimeters.  
Oscillations into bulk neutrino will now result. The
profile of the oscillations in the present case
is quite different from the case of models with global
$B-L$. Now, the mass eigenstates obey  the  characteristic equation
$\lambda_n = \pi \kappa^2 v_R^2 R^2 \ \cot(\pi \lambda_n)$. 
For the weak eigenstate we found
$\nu_e= \cos\theta ~\nu_0 + \sin\theta ~\tilde\nu_0$,
where $\tan\theta =\frac{hv}{\kappa v_R}$; and 
$\tilde\nu_0$ is given in terms of the mass eigenstates as 
 $\tilde\nu_0(t) = \sum{1 \over \eta_n} \nu_n$,
with the mixing factors given by  
$\eta_n^2 =2(\lambda_n^2/\zeta^2)(\lambda_n^2 + f(\zeta))$, with 
$\zeta=\sqrt{2} \kappa v_R R$ and $f(\zeta)$ as before. 
Thus, in 
this case the KK contributions enter in
$\nu_e$ trough the universal mixing
angle $\theta$, in contrast with Eq. (\ref{nul}). 
Now, the survival probability after the neutrino traverses a distance L in
vacuum  reads
$P_{surv}(L)= \cos\theta^4 + 
\sin\theta^4 ~|\la\tilde\nu_0|\tilde\nu_0(L)\ra|^2
+ 2 ~\cos\theta^2 \sin\theta^2 ~ Re \la \tilde\nu_0|\tilde\nu_0(L)\ra$.
The averaged probability has the form  
$\overline{P_{surv}} = \cos\theta^4 + {2\over 3} \sin\theta^4$, which is
smaller than the two neutrino case with the same mixing angle, though,
for small mixing it approaches the former  result. 

Further analysis extending the present model to three brane generations
was presented in Ref.~\cite{mp2}. There, seesaw Majorana terms were
included, and a single bulk neutrino  plays the role of a sterile
neutrino with its lightness associated to the  geometry of the extra
dimension. The possible scenario that explain the neutrino anomalies could
be as follows: solar data is given by $\nu_e\rightarrow\nu_B$
oscillations and matter effects, this implies a submillimeter radius. 
Atmospheric data, is
provided by the usual $\nu_\mu\rightarrow\nu_\tau$ oscillations thanks to
a natural  decoupling of this sector from the KK modes.  Finally, LSND is
explained by a small $\nu_e-\nu_\mu$  mixing. 

\section{Standard Model in extra dimensions}
\seq

Considering the  possibility  that the SM particles propagate on the extra
dimensions drives the models back to the former KK theories, 
nevertheless, besides the new possibility of having 
large extra dimensions,  the
fact that some particles could be still attached to the branes 
makes this scenario quite different from the
former one. Before considering any possible case  one must notice that the
conservation of the charges, that is, the consistency of the local gauge
symmetry, implies that the first natural candidates to propagate in the
bulk are the gauge bosons. Once they are promoted to be bulk fields
 we will think on the SM fields
as the zero modes of higher dimensional fields.  However,  as there is not
experimental evidence of light copies of $Z$, $W$, etc., we lead to the
conclusion that this models can not have a too large extra dimension. The
current experimental data provide lower bounds for the size of $R$ just at
the TeV scale, suggesting that these extra dimensions may show up in the
near future at the colliders.   The whole scenario could be as follows:
There are several extra dimensions. The SM particles are free to propagate
within one (or more) $p$-brane(s),  where $p>3$ and where  the largest
extra compact dimensions are about TeV's,  while gravity lives in a
higher dimensional bulk with some large (millimetric) extra dimensions. 
It is worth mentioning that this  scenario could  in fact be realized
from string theory.
To simplify the discussion of this models we will follow Ref.~\cite{sm}, 
although we recommend the reader to see also the important early works,
some of which are given in  Refs.~\cite{quiros,earlier,mirabelli,ddg,kaku}.

\subsection{Theoretical setup}

Let us consider again  five dimensions, with an orbifolded 
fifth dimension.
Then, let us assume that  the SM
gauge  fields live in the bulk, while  fermions, $\psi$, and  Higgs
doublets $H_i$ can either live in the bulk or  on the  $y=0$ brane. The
analysis will follow for two Higgs doublets to make a possible 
extension to
supersymmetry obvious. The case with only one scalar doublet is
straightforward. The lagrangian reads 
 \begin{eqnarray}
 {\cal L}_{5D}&=&-\frac{1}{4}
 F_{MN}^2+
 \sum_{i}
 \left[(1-\varepsilon^{H_i})
 |D_M H_{i}|^2+(1-\varepsilon^{\psi_i})
 i\bar \psi_{i}\Gamma^M D_M \psi_{i}\right]\nonumber\\
 &&+\sum_{i}
 \left[\varepsilon^{H_i}|D_\mu H_{i}|^2+
 \varepsilon^{\psi_i}i\bar \psi_{i}\sigma^\mu D_\mu \psi_{i}
 \right]
 \delta(y)\, ,
 \end{eqnarray}
where  $\varepsilon^{F}=1\ (0)$ when the $F$-field lives on the boundary 
(bulk); $D_M=\partial_M+ig_5V_M$; $V_M = V_M^a T^a$, with $T^a$ the
group generators 
 and $g_5$  the 5D gauge coupling. Clearly a $g'_5$ should be introduced
 for $U(1)_Y$. Notice that the effective four
 dimensional couplings obeys $g= g_5/\sqrt{\pi R}$, thus the gauge sector
 should be strongly coupled on the bulk.
Gauge and Higgs bosons living in the 5D bulk 
are  assumed to be even under the ${\cal Z}_2$ symmetry.
We will choose the even assignment for the
$\psi_L$ ($\psi_R$) components of 
fermions $\psi$ which are doublets (singlets)
under SU(2)$_L$, this is in order to recover the low energy SM spectra. 

At intermediate energies, below the compactification scale, $M_c$, 
the impact of this theory on standard (boson exchange)  processes 
may be studied
by integrating out the KK modes, and summing up the diagrams as we did for
Eq. (\ref{msum}). Thus, let us introduce the useful small parameter
 \begin{equation}
 X=\sum_{n=1}^\infty \frac{2}{n^2}\, \frac{m_Z^2}{M_c^2}=\frac{\pi^2}{3}
 \frac{m_Z^2}{M_c^2}\, ,
 \label{equis}
 \end{equation}
and do all the analysis at first order on $X$. Also,  it is useful to
introduce the  effective mixing angle
$s^2_\alpha=\sin^2\alpha = 
\varepsilon^{H_2}\,s^2_\beta+\varepsilon^{H_1}\,c^2_\beta $
where, as usual, $\tan\beta=\langle H_2\rangle/\langle H_1\rangle$,
with $v^2\equiv\langle H_1\rangle^2+ \langle H_2\rangle^2\simeq
(174~GeV)^2$.
In these terms the charged sector in the four dimensional effective theory 
and in the unitary gauge has the form~\cite{sm}
 \begin{equation}
 {\mathcal L}^{cc}_{eff}=\frac{1}{2} M^{2}_W
 W\cdot W -g W\cdot\left[J-s^2_\alpha\, c_\theta^2
 \,X\,  J^{KK}\right]
 -\frac{g^2}{2\, m_Z^2}\,X\,J^{KK}\cdot J^{KK}\, ,
 \label{leff1}
 \end{equation}
where $M_W^2=\left[1-s^4_\alpha c^2_\theta X\right]\, m_W^2\,$; being 
$m^{2}_W=g^2v^2/2$ and $\theta$ the usual electroweak mixing  
angle. $J_\mu$ and $J_\mu^{KK}$ are the fermion currents of the zero
and excited modes respectively. The last term in (\ref{leff1})
is an effective four
points interaction induced by the exchange of the heavy $W^{KK}$. Note that
the tree level mass of the $W$ receive a contribution from its 
mixing with
the excited modes   if the Higgs is a brane field ($\varepsilon^H=1$).
This lagrangian induce a tree level correction to the Fermi constant from
the $\mu$ decay, that  reads
 \begin{equation}
 \frac{G_F}{\sqrt{2}}=\frac{g^2}{8 M_W^2}
 \left[1+\lL c_{2\alpha} c^2_\theta 
  X \right]\, . \label{fermiun}
 \end{equation}

On the other hand, for  the neutral currents we get at the same limit
 \begin{eqnarray}
 {\mathcal L}^{nc}_{eff}&=&\frac{1}{2} M_Z^2\, Z\cdot Z 
 -\frac{e}{s_\theta\, c_\theta}\, Z\cdot\left[J_Z-\,
 s^2_\alpha X\,  J_Z^{KK}
 \right]-e A\cdot J_{em}
 \nonumber\\
 &-&\frac{1}{2\,M_Z^2}\,\frac{e^2}{s^2_\theta\, c^2_\theta}\,X
 J_Z^{KK}\cdot J_Z^{KK}
 -\frac{e^2}{2\, M_Z^2}\,X\, J_{em}^{KK}\cdot J_{em}^{KK}\, ,
 \end{eqnarray}
where $M_Z^2= \left[1-s^4_\alpha X \right]m_Z^2$. As before the $J$'s 
represent
the usual four dimensional currents and $J^{KK}$ that corresponding to the
matter KK excitations, if they exist. Note that the zero mode of the photon
($A$) remain massless. From the  $W$ and $Z$ masses combined with Eq.
(\ref{fermiun}), one may relates the weak mixing angle to the $G_F$ as 
 \begin{equation}
 \label{relacion1}
 s^2_\theta\, c^2_\theta=\frac{\pi\alpha}{\sqrt{2}\, G_F\, M_Z^2}\,
 \left(1+\Delta\right)\, ,
 \end{equation}
where the parameter
$\Delta=\left[\lL c_{2\alpha}c^2_\theta-s^4_\alpha s^2_\theta \right]\, X$;
and $\alpha$ is the fine structure constant.  

Another  ingredient that may be reinstalled on the theory is supersymmetry.
Although it is not necessary to be considered, it is an interesting
extension. After all, it seems plausible to exist  if the high energy
theory is  string theory. If the theory  is supersymmetric,  the bulk
fields come in  $N=2$  supermultiplets~\cite{mirabelli,quiros}. The
on-shell field content of the gauge supermultiplet is
${V}=(V_\mu,V_5,\lambda^i, \Sigma)$ where $\lambda^i\ (i=1,2)$ is a
simplectic Majorana spinor  and $\Sigma$ a real scalar in the adjoint
representation; $(V_\mu,\lambda^1)$ is  even under ${\cal Z}_2$ and 
$(V_5,\Sigma,\lambda^2)$  is odd.  Matter and Higgs fields are arranged in
$N=2$ hypermultiplets that consist of  chiral and antichiral $N=1$
supermultiplets. The chiral $N=1$ supermultiplets are even under 
${\cal Z}_2$ 
and contain  massless states. These will correspond to  the SM fermions
and Higgses. 

Supersymmetry must be broken by some mechanisms that  gives masses to all
the superpartners which we may assume are  of order $M_c$~\cite{quiros}. 
For some possible mechanism see Ref.~\cite{mirabelli}. 
In contrast with the case of four dimensional susy, where no extra effects
appear at tree level after integrating out the superpartners, in the
present case integrating out the  scalar field $\Sigma$ induces a
tree-level contribution to  $M_W$ and $\Delta$ parameters.  In
Ref.~\cite{sm} they were calculated to be given by 
$M_W^2=\left[1-\hone\htwo\, s_{2\beta}^2\, c^2_\theta X\right]\, m_W^2$; 
and  $\Delta=\left[\lL c_{2\alpha}\, c^2_\theta-s^4_\alpha +\hone \htwo \,
s^2_{2\beta}\, c^2_\theta\right]\, X$ respectively. The form of the low
energy lagrangian remains.

\subsection{Experimental constrains} 

There are two important effects of gauge KK boson states on collider
experiments.  (i) Mixing effects  and  (ii) real production of KK modes. 

First, the mixings between the zero  and the KK modes of gauge bosons 
modify the SM observables, affecting the Electroweak precision
tests~\cite{coll2,sm,review1}.  
Lets consider for instance a specific non supersymmetric 
model that fixes fermions and one
Higgs doublet to the boundary, while gauge bosons and another Higgs
doublet propagates on the bulk. That fixes $s_\alpha = s_\beta$ in our
expressions above. Then the model has two more parameters than the SM, 
given by $s_\beta$ and $X$, or something equivalent.
All observables will be expressed explicitly or implicitly 
in terms of these and the  
usual SM parameters. For example, Rizzo and Wells in
Ref.~\cite{coll2}, introduce the new parameter
$V= {M_W^2\over m_Z^2} X$ and the effective interaction  couplings 
$g_W = g[1-s_\beta^2 V]$ and 
$g_Z = g[1-s_\beta^2 V/c_\theta^2 ]$ to  get
 \[
 G_F(\mu~{\rm decay}) =  \frac{\sqrt{2}g_W^2}{8M_W^2}[1+V],\qquad\quad
 \Gamma(Z\to f\bar f) =  \frac{N_c M_Z}{12\pi}
  \left( \frac{g_Z}{2c_\theta}\right)^2 
   \left[ v_f^2 + a_f^2 \right], 
 \]
 \[
 Q_W  = \frac{1}{M^2_Z}\left\{ \frac{g^2 (1-s^2_\beta V/c_\theta^2 )^2}
  {c^2_\theta} + \frac{g^2V}{c_\theta^4}\right\} 
  a_e \left[ v_u (2Z+N)+v_d(2N+Z)\right] , 
  \]
 \be
 R  =  \frac{\sigma^\nu_{\rm NC} -\sigma^{\bar \nu}_{\rm NC}}
 {\sigma^\nu_{\rm CC} -\sigma^{\bar \nu}_{\rm CC}} =
 \left[ \frac{g^2_Z}{c_\theta^2M_Z^2}
          +\frac{g^2 V}{c_\theta^4 M_Z^2}\right]
 \left[ \frac{g^2_W}{M^2_W}+\frac{g^2V}{M^2_W}\right]^{-1} 
 \left( \frac{1}{2}-s^2_\theta\right) ,
 \ee
\[
 A_f =  \frac{2v_fa_f}{v_f^2+a_f^2}, \qquad\quad
 \sin^2\theta^{\rm eff}_W = x + \frac{x(1-x)}{1-2x}V
    \left[ c^4_\beta-\frac{s^4_\beta}{1-x}\right],
 \]
 \[
 M^2_W=M^2_Z(1-x)\left\{ 1+V\left[ 1-2s^2_\beta-
        \frac{c_\beta^4(1-x)-s^4_\beta}{1-2x}\right]\right\},
 \]
where $Q_W$ is a measure of atomic parity violation, $x$ is the solution
to the equation
$x(1-x)=\pi\alpha/\sqrt{2}G_F m^2_Z$;
$v_f\equiv T_{3f}-2Q_f s^2_\theta$ and $a_f\equiv T_{3f}$.
Overall, the limit on $M_c$
using the precision data
measurements~\cite{pdg}  is just about $M_c \gsim 3.3 - 3.8$
TeV~\cite{review1}. 

Future colliders may be able to observe  resonances due to  KK modes  if
the compactification scale turns out to be on the TeV range. This needs a
collider energy $\sqrt{s}\gsim M_c$. In hadron colliders (TEVATRON, LHC) 
the KK excitations might be directly produced in Drell-Yang processes 
$pp(p\bar p) \rightarrow \ell^-\ell^+X$ where the lepton pairs 
($\ell=e,\mu,\tau$) are produced via the subprocess $q\bar q\rightarrow
\ell^+\ell^+X$.  This is the more useful mode to search for 
$Z^{(n)}/\gamma^{(n)}$ even $W^{(n)}$. 
Current search for $Z'$ on this channels (CDF)
impose $M_c> 510~GeV$. Future bounds could be raised up to $650~GeV$ in
TEVATRON and $4.5~TeV$ in LHC, which with  100 $fb^{-1}$ of luminosity can
discover modes up to $M_c\approx 6~TeV$. 

Deviations on the cross sections due to virtual exchange of KK modes may
be observed in both, hadron and lepton colliders.  With a  $20~fb^{-1}$
of luminosity, 
TEVATRONII may observe signals up to $M_c \approx
1.3~TeV$. LEPII with a maximal luminosity of 200 $fb^{-1}$ could impose
the bound at $1.9~TeV$, while NLC may go up to 13 $TeV$, which slightly
improve the bounds coming from precision test.

\section{Gauge coupling unification}
\seq

Once we have assumed a low  fundamental scale for quantum gravity, the
natural question is whether the former picture of a Gran Unified
Theory~\cite{guts} should be abandoned and with it a possible gauge theory
understanding of the quark lepton symmetry  and gauge hierarchy. On the
other hand, if string theory were the right theory above $M$  an unique
fundamental coupling constant would be expect, while the SM contains three
gauge coupling constants. Then, it
seems clear that, in any case, a sort of low energy  gauge coupling
unification is required. As pointed out in Ref.~\cite{ddg} and later
explored in~\cite{kaku,uni,uni2,unimore,kubo},  if the SM particles live
in higher dimensions, as in the model discused above, such a low GUT scale
could be realized.

For comparison let us mention how one leds to gauge unification in four
dimensions. Key ingredient in our discussion are the renormalization
group  equations (RGE) for the gauge coupling parameters that at one loop,
in the $\overline{MS}$ scheme,
read
\be
\frac{d\alpha_i}{dt}=\frac{1}{2\pi} b_i \alpha^2_i
\label{rge}
\ee
where $t=ln \mu$. $\alpha_i = g_i^2/4\pi$; $i=1,2,3$, are the coupling
constants of the  SM factor groups $U(1)_Y$, $SU(2)_L$ and $SU(3)_c$
respectively.  The coefficient $b_i$ receives contributions from the 
gauge part and the matter including Higgs field  and its  completely
determinated by 
 $4\pi b_i = {11\over 3} C_i(vectors) - {2\over 3}C_i(fermions)
 -{1\over 3}C_i(scalars)$,
where $C_i(\cdots)$ is the index of the representation to which the
$(\cdots)$ particles are assigned, and where we are considering Weyl
fermion and complex scalar fields. Fixing the normalization of the $U(1)$
generator as in the $SU(5)$ model, we get for the SM  $(b_1,b_2,b_3)
=(41/10,-19/6,-7)$ and for the Minimal Supersymmetric SM (MSSM)
$(33/5,1,-3)$.  Using Eq. (\ref{rge}) to extrapolate the values measured
at the $M_Z$ scale~\cite{pdg}: $\alpha^{-1}_1(M_Z)=58.97\pm .05$;
$\alpha^{-1}_2(M_Z)=29.61\pm .05$; and  $\alpha^{-1}_3(M_Z)=8.47\pm .22$,
(where we have taken for the strong coupling constant the global average),
one finds that only in the MSSM  the three couplings merge together at the
scale $M_{GUT}\sim  10^{16}$ GeV. This high scale naturally explains the
long live of the proton and  
in the minimal $SO(10)$ framework one gets a very
compelling and predictive scenario~\cite{pati}.

A different possibility for unification   that does not involve
supersymmetry is the existence of an intermediate left-right
model~\cite{guts} that breaks down to the SM symmetry at $10^{11-13}~GeV$.
It is worth mentioning that a non canonical
normalization of the gauge coupling may, however, substantially change the
above figures, predicting a different unification scale.
Such a different normalization may arise either in some no minimal 
unified
models, or in string theories
where the SM group
factors are realized on non trivial Kac-Moody
levels~\cite{ponce,kdienes}. Such  scenarios are in general
more complicated than the minimal $SU(5)$ or $SO(10)$ models since they
introduce new exotic particles.

\subsection{Power law  running}

It is clear that the presence of KK
excitations will affect the evolution of couplings in gauge theories and
may alter the whole picture of unification of couplings. 
This question was first studied  by Dienes, Dudas and
Gherghetta (DDG)\cite{ddg} on the base of the effective
theory approach at one loop level. They found that 
above the compactification scale  $M_c$ one gets
 \be
 \alpha_i^{-1}(M_c) = \alpha_i^{-1}(\Lambda) + {b_i - \tilde{b}_i\over
2\pi}
 \ln\left( {\Lambda\over M_c}\right) + {\tilde{b}_i\over 4\pi}\
 \int_{r\Lambda^{-2}}^{rM_c^{-2}}\! {dt\over t}
 \left\{ \vartheta_3\left({it\over \pi R^2}\right)\right\}^\delta ,
 \label{exact}
 \ee
with $\Lambda$ as the ultraviolet cut-off and  
$\delta$  the number of extra
dimensions.
The Jacobi theta function
 $\vartheta(\tau) = \sum_{-\infty}^{\infty} e^{i\pi \tau n^2}$
reflects the sum over the complete tower. Here
$b_i$ are the beta functions of the theory below $M_c$,  
and $\tilde{b}_i$ are the contribution to the beta functions 
of the KK states at each excitation
level. 
The numerical factor $r$ depends on the renormalization scheme.
For practical purposes, we may approximate the above result by 
decoupling all the
excited states with masses above $\Lambda$, and assuming that the
number of KK states below certain energy $\mu$ between $M_c$ and
$\Lambda$ is well approximated by the volume of a $\delta$-dimensional
sphere of radius ${\mu\over M_c}$ given by
 $N (\mu,M_c) = X_\delta
 \left({\mu\over M_c}\right)^\delta$; 
with $X_\delta = \pi^{\delta /2}/\Gamma(1 +\delta /2)$.
The result is a power law behaviour of the gauge coupling
constants~\cite{taylor}:
 \be
 \alpha_i^{-1}(\mu) = \alpha_i^{-1}(M_c) - {b_i - \tilde{b}_i\over 2\pi}
 \ln\left( {\mu\over M_c}\right) - {\tilde{b}_i\over 2\pi}\cdot
 {X_\delta\over\delta}\left[ \left({\mu\over M_c}\right)^\delta - 1\right],
 \label{ddgpl}
 \ee 
which  accelerates the meeting of the $\alpha_i$'s.
In the MSSM the energy range between $M_c$ and $\Lambda$ --identified as
the unification (string) scale M-- is relatively small due to the steep
behaviour in the evolution of the couplings~\cite{ddg,uni}. For instance,
for a single extra dimension the ratio $\Lambda/M_c$ has an upper limit of
the order of 30, and it substantially decreases for larger $\delta$.

This same relation can be understood on the basis of a 
step by step approximation~\cite{uni2}  as
follows. We take the SM gauge couplings and extrapolate their values up to
$M_c$ then we add to the beta functions the contribution of the first KK
levels, then we run the couplings upwards up to
just below the next consecutive level where
we stop and add the next KK contributions, and so on, 
until the energy $\mu$.
Despite the complexity of the spectra, the degeneracy of each level 
is always computable and performing a level by level approach of the gauge
coupling running is  possible.
Above the  $N$-th level the running  receives
contributions from $b_i$ and of
all the KK excited states in the
levels below, in total $f_\delta(N) = \sum_{n=1}^N g_\delta(n)$,
where $g_\delta(n)$ represent the total degeneracy of the level $n$. 
Running for all the first $N$ levels  leads to
 \be
 \alpha_i^{-1}(\mu) = \alpha_i^{-1}(M_c) - {b_i\over 2\pi}
 \ln\left( {\mu\over M_c}\right) - {\tilde{b}_i\over 2\pi} \left[
 f_\delta(N)\ln\left(\mu\over M_c\right) - 
 {1\over 2}\sum_{n=1}^N g_\delta(n)\ln n  \right].
 \label{hdlog}
 \ee
A numerical comparison of this expression with the
power law running 
shows the accuracy of that approximation. Indeed, in the
continuous limit the last relation  reduces into Eq. (\ref{ddgpl}).
Thus, gauge coupling unification may now  happen at TeV
scales~\cite{ddg}.
Next, we will discuss how accurated this unification is.

\subsection{One step unification}

Many features of unification can be studied
without bothering about the detailed subtleties of the running. 
Consider the generic form for the 
evolution equation 
 \be
 \alpha_i^{-1}(M_Z) = \alpha^{-1} + {b_i\over 2\pi}
 \ln \left( {M\over M_Z}\right) + {\tilde{b}_i\over 2\pi} 
 F_\delta\left( {M\over M_c}\right), 
 \label{rgef}
 \ee 
where we have changed $\Lambda$ to $M$ to keep our former notation.  
Above, $\alpha$ is the unified coupling and $F_\delta$ is  given by
the expression between parenthesis in Eq. (\ref{hdlog}) or its
correspondent limit in Eq. (\ref{ddgpl}).
Note that the information that comes from the bulk is being
separated into two independent parts: all the structure of the KK spectra 
$M_c$ and $M$ are  completely embedded into the
$F_\delta$ function, and  their contribution is actually
model independent. The only (gauge)  model dependence comes in the beta
functions, $\tilde{b}_i$. 
Indeed, Eq. (\ref{rgef}) is similar to that of  the two step unification
model where a new gauge symmetry appears at an intermediate energy scale.
Such models are very constrained by the one step unification  in the MSSM.
The argument goes as follows: let us define the vectors: 
${\bf b}= (b_1,b_2,b_3)$; 
$\tilde{\bf b}= (\tilde{b}_1,\tilde{b}_2,\tilde{b}_3)$; 
${\bf a} = (\alpha_1^{-1}(M_Z),\alpha_2^{-1}(M_Z),\alpha_3^{-1}(M_Z))$ and 
${\bf u} = (1,1,1)$, and construct the unification barometer~\cite{uni2}
$\Delta\alpha\equiv ({\bf u}\times {\bf b})\cdot {\bf a}$.
For single step unification models the unification
condition amounts to the condition $\Delta\alpha=0$.
As a matter of  fact, 
for the SM $\Delta\alpha=41.13\pm 0.655$, while for the MSSM
$\Delta\alpha = 0.928\pm 0.517$, leading to unification within two
standard deviations. In this notation 
Eq. (\ref{rgef}) leads to 
 \be
 \Delta\alpha = 
 [({\bf u}\times {\bf b})\cdot \tilde{\bf b}]\ {1\over 2 \pi}F_\delta.
 \label{dalpha}
 \ee
 Therefore, for the MSSM, 
we get the constrain~\cite{mohapatra}
 \be
 (7\tilde{b}_3 - 12\tilde{b}_2 + 5 \tilde{b}_1)F_\delta= 0.
 \ee
There are two solutions to the this equation: 
(a) $F_\delta(M/M_c)=0$, which means $M = M_c$,
bringing  us back to the MSSM by pushing up the compactification scale to
the unification scale.
(b) Assume that the beta coefficients $\tilde{b}$ conspire to 
eliminate
the term between brackets: 
$(7\tilde{b}_3 - 12\tilde{b}_2 + 5 \tilde{b}_1) = 0$,
or equivalently~\cite{ddg}
 \be
 {B_{12}\over B_{13} } = {B_{13}\over B_{23} } = 1; \qquad \mbox{where}\qquad
 B_{ij} = {\tilde{b}_i- \tilde{b}_j\over b_i- b_j} .
 \ee
The immediate consequence of last possibility is the indeterminacy  of
$F_\delta$, which means that we may put $M_c$ as a free parameter in the
theory. For instance we could choose $M_c\sim$ 10 TeV to maximize the
phenomenological impact of such models.  It is compelling to stress that
this conclusion is independent of the explicit form of $F_\delta$.
Nevertheless, the minimal model where all the  MSSM particles propagate on
the bulk does not satisfy that constrain~\cite{ddg,uni}. Indeed, in this
case we have $(7\tilde{b}_3 - 12\tilde{b}_2 + 5 \tilde{b}_1) = -3$, which
implies a  higher prediction for $\alpha_s$ at low $M_c$. As lower the
compactification scale, as higher the prediction for $\alpha_s$. 
However,  as
discussed  in Ref.~\cite{uni} there are some  scenarios where the
MSSM fields are distributed in a nontrivial  way among the bulk and the
boundaries which lead to unification. There is also the  obvious
possibility of adding  matter to the MSSM to correct the accuracy on
$\alpha_s$. 

The SM case has similar complications. 
Now Eq. (\ref{rgef}) turns out to be a system of three equation with three
variables, then, within the experimental accuracy on $\alpha_i$, 
specific predictions for $M$, $M_c$ and $\alpha$ will arise. 
As $\Delta\alpha\neq 0$, the
above constrain does not aply, instead the matter content should
satisfy the consistency conditions~\cite{uni2}
 \be 
 Sign(\Delta\alpha)  
 = Sign[({\bf u}\times {\bf b})\cdot \tilde{\bf b}]
 = - Sign(\tilde{\Delta}\alpha) \ ;
 \label{cons2}
 \ee
where $\tilde{\Delta}\alpha\equiv({\bf u}\times\tilde{\bf b})\cdot{\bf a}$.
However, in the minimal model where all SM fields are assumed to have KK
excitations  one gets 
$\tilde{\Delta}\alpha = 38.973\pm 0.625$; and 
$({\bf u}\times {\bf b})\cdot \tilde{\bf b}^{SM} = 1/15$. 
Hence, the constraint (\ref{cons2}) is not
fulfilled and unification does not occur. 
Extra matter could of course improve this 
situation~\cite{ddg,uni,uni2}. Models with non canonical normalization may
also modify this conclusion~\cite{uni2}. A particularly
interesting outcome in this case is that
there are some cases where, without introducing extra matter at the SM
level, the unification scale comes out to be
around $10^{11}$ GeV (for instance $SU(5)\times SU(5)$,
$[SU(3)]^4$ and $[SU(6)]^4$). 
These models fit nicely into the new intermediate string scale models
recently proposed in~\cite{quevedo}, and also with the expected scale in
models with local $B-L$ symmetry. High order
corrections has been considered in Ref.~\cite{unimore}. 
The analysis for
the  running of other coupling constants could be found
in~\cite{ddg,kubo}. Two step models were also studied in~\cite{uni2}.

Now well,  with a TeV unification scale 
a extremely fast proton decay mediated by new gauge interactions may
occur. 
There are two possible solutions to this problem. The obvious one
is invoking an unified group that keeps the proton stable. 
A less trivial possibility was suggested
in~\cite{ddg}. If the gauge bosons that mediate proton decay are odd under
the ${\cal Z}_2$ symmetry of the orbifold, then their coupling to the
quarks (fixed at the brane) is forbidden, and the proton remains stable.
On the context of string theories it may also happen that gauge coupling
unification occurs  without the appearance of 
any extra gauge symmetry at the string scale. 
We should note, however, that
this last mechanism does not 
remove the danger of having a  fast proton decay induced
by high order operators.

\section{Splitting  wave functions on  thick walls}
\seq

As we already mentioned  some mechanism is needed in this theories  to
forbid dangerous higher dimension operators  which lead to proton decay,
large neutrino masses, etc., since they are now
suppressed just by $M$. Without the knowledge of the theory above $M$  it
is difficult just to assume that  such operators are not being induced. 
One might
of course invoke global symmetries again. However, an interesting 
mechanism that explain how proton decay could get  suppressed at the
proper level appeared in~\cite{sch1}.  It relays on the idea that the
branes are being formed from an effective mechanism that traps the SM
particles in it, resulting in a  wall with thickness $L\sim M^{-1}$, where
the  fermions are stuck at different points.  Then, fermion-fermion
couplings get suppressed due to the exponentially small  overlaps of their
wave functions. This provides a framework for understanding both the
fermion mass hierarchy and proton stability without imposing extra 
symmetries, but rather in terms of a higher dimensional
geography~\cite{sch2}.   Note that the dimension where the gauge fields
propagate does not need to be orthogonal to the millimetric
dimensions, but  gauge  fields may be restricted to live in a smaller
part of that extra  dimensions.  Here we briefly summarize those ideas.

\subsection{Localizing wave functions on the brane}

Let us start by assuming that the translational invariance along the fith
dimension is being broken
by  a bulk scalar field $\Phi$
which develops a spatially varying expectation value $\la\Phi\ra(y)$. We
assume that this expectation
value have the shape of a domain wall transverse to the extra
dimension and is centered at $y=0$. With this background 
a bulk fermion will have a zero mode that is stuk at the zero of 
$\la\Phi(y)\ra$. To see this let us consider the action
 \be 
 S= \int d^4x\ {\rm d}y\,
 \overline\Psi\left[i\, \Gamma^M\!\partial_M +
 \la\Phi\ra(y)\right] \Psi ,
\label{single5}
 \ee 
in the chiral basis, as before. By introducing the expansions
 \be
  \Psi_L(x,y) = \sum_n f_n(y)\psi_{nL}(x);\qquad \mbox{ and } \qquad
 \Psi_R(x,y) = \sum_n g_n(y)\psi_{nR}(x);
 \ee
where $\psi_{n}$ are four dimensional spinors, we get for the 
$y$-dependent functions $f_n$ and $g_n$ the equations
 \be
 \left( \partial_5 +\la\Phi\ra\right) f_n  + \mu_n g_n= 0;
 \qquad \mbox{ and } \qquad
 \left(-\partial_5+\la\Phi\ra \right) g_n + \mu_n f_n= 0.
 \ee
Therefore, the zero modes have the profiles~\cite{sch1}
 \be
 f_0(y)\sim \exp\left[-\int^{y}_0\! ds \la\Phi\ra(s) \right]
 \qquad \mbox{ and }\qquad
 g_0(y) \sim \exp\left[\int^{y}_0 ds \la\Phi\ra(s)\right];
 \ee
up to normalization factors. Notice that when the extra space is supposed
to be finite, both modes are normalizable. For the special choice
$\la\Phi\ra(y)= 2\mu^2y$, we get $f_0$ centered at $y=0$ with the gaussian
form
 \be 
 f_0(y)= \frac{\mu^{1/2}}{(\pi/2)^{1/4}}\ \exp\left[{-\mu^2{y}^2}\right] .
 \ee
The other mode has been
projected out from our brane by being pushed away to the end of the space.
Thus, our theory in the wall is a chiral theory. 
Notice that a negative coupling among $\Psi$ and $\phi$ will instead
project out the left handed part.

The generalization of this technique to the case of several fermions is
straightforward. The action (\ref{single5}) is generalized to
 \be 
 S = \int\! d^5x\, 
 \sum_{i,j}\bar\Psi_i[i\,\Gamma^M\partial_M + \lambda\la\Phi\ra-m]_{ij} \Psi_j\ ,
 \label{multi5}
 \ee
where general Yukawa couplings $\lambda$ 
and other possible five dimensional masses $m_{ij}$
have been considered. For simplicity we will assume both terms diagonal.
The effect of these new parameters is a shifting of the wave functions,
which now are centered around the zeros of $\lambda_i\la\Phi\ra-m_i$.
Taking $\lambda_i =1$ with the same profile for the vacuum leads to
gaussian distributions centered at $y_i = m_i/2\mu^2$. Thus, at low
energies, the above action will describe a set of non interacting four
dimensional chiral fermions localized at different positions in the fifth
dimension.

Localization of gauge and Higgs bosons  needs extra assumptions. The
explanation of this phenomena is close related with the actual way the
brane was formed. A field-theoretic mechanism for localizing gauge fields
was proposed by Dvali and Shifman and was later extended and applied  in
\cite{dvali}. There, the idea is to arrange for the gauge group to confine
outside the wall; the flux lines of any electric sources turned on inside
the wall will then be repelled by the confining regions outside and forced
to propagate only inside the wall. This traps a massless gauge field on
the wall. Since the gauge field is prevented to enter the confined region,
the thickness $L$ of the wall acts effectively as the size of the extra 
dimension in which the gauge fields can propagate. In this
picture, the gauge couplings will exhibit power law running
above the scale $L^{-1}$.

\subsection{Fermion mass hierarchies and proton decay}

Let us consider the Yukawa coupling among the Higgs field and the leptons:
$\kappa H  L^{T}E^c$; where the massless zero
mode  $l$ from $L$ 
is localized at $y=0$ while $e$ from 
$E^c$ is localized at $y=r$. Let us also assume that the Higgs zero mode is
delocalized inside the wall. Then the zero modes term of this coupling will
generate the effective Yukawa action
 \be 
  S_{Yuk} = \int\! d^4 x\, \kappa\, h(x) l(x) e^c(x)\ 
  \int\! dy\ \phi_l(y)\ \phi_{e^c}(y)\ ,
 \ee
where $\phi_l$ and $\phi_{e^c}$ represent the gaussian profile of the
fermionic modes. Last integral gives the overlap of the wave functions,
which is exponentially suppressed~\cite{sch1} as
 \be 
  \int\! dy\ \phi_l(y)\ \phi_{e^c}(y)\ = e^{-\mu^2r^2/2}.
 \ee
This is a generic feature of this models. The effective coupling of any
two fermion fields is exponentially suppressed in terms of their
separation in the extra space. Thus, the  explanation for the mass
hierarchies becomes a problem of the cartography on the extra
dimension. A more detailed analysis was presented in~\cite{sch2}.

Let us now show how  a fast proton decay 
is evaded in these models.  
Assume, for instance,   that all quark fields are localized at 
$y=0$ whereas the leptons are at $y=r$. 
Then, lets consider the following baryon
and lepton number violating operator
\be
S \sim \int d^5x\,  \frac{(Q^{T} C_5 L)^{\dagger} (U^{c T}
C_5 D^c)}{M^3} .
\ee
In the four dimensional effective theory, once we have 
introduced the zero mode
wave functions, we get the suppressed action~\cite{sch1}
\be
S \sim \int d^4x\,\, \lambda \times \frac{(q l)^{\dagger}
(u^c d^c)}{M^2}
\ee
where $\lambda \sim \int dy\, \left[ e^{-\mu^2y^2}\right]^3
e^{-\mu^2(y-r)^2}\sim e^{-3/4 \mu^2r^2}$.
Then,  for a separation of $\mu r =10$ we obtain $\lambda\sim10^{-33}$
which renders these operators completely safe even for $M\sim 1$ TeV.
Therefore, we may 
imagine a picture where quarks and leptons are localized near
opposite ends of the wall so that $r \sim L$.
This mechanism, however, does not work for suppressing the another
dangerous operator
$(LH)^2/M$ responsible of a  large neutrino mass.

\section{Randall Sundrum model and other current trends}
\seq

To close our present discussion allow us to mention another
important direction of research in this area. It was
motivated by a seminal work of Randall and Sundrum~\cite{rs}, who proposed
a drastic change on our present point of view of the 
way the bulk enters  in the explanation of the hierarchy problem. 
Here we summarize some aspects of this model and some further
trends~\cite{gw,csaki,rsth,rizzo,grossman,cc,binetruy,cline,cosmo} that
may give a rough idea of the way this area is going.

\subsection{Mass Hierarchy from a Small Extra Dimension}

Lets consider the following setup. A five dimensional space with an
orbifolded fifth
dimension. Consider two branes at the fixed points
$y=0,\pi$; with tensions $\sigma$ and $-\sigma$ respectively. Assume
that the bulk has a cosmological constant $\Lambda$. Contrary
to our previous philosophy, here let us  assume that all parameters are of the
order of the Planck scale. Moreover, we will 
not longer assume that the bulk
has a flat metric, instead we will consider 
a non factorizable geometry induced
by the (explicitly) broken translational invariance. Thus, the more
general metric that respects four dimensional Poincare invariance on the
brane  has the form: 
 \be 
  ds^2 = G_{AB} dx^A dx^B = 
  e^{-2\beta(y)}g_{\mu\nu}(x)dx^\mu dx^\nu - r^2dy^2.
 \ee
By solving the Einstein equations with this ansatz for the metric, it turns
out that $\sigma$ and $\Lambda$ need to be related by the (fine tunning)
condition
 \be 
 \Lambda  = - {\sigma^2\over 6 M^3};
 \ee
that is equivalent to the exact cancellation of the effective four
dimensional cosmological
constant. On the other hand, one gets
 $\beta(y) = kr|y|$, where  $k^2 = {\Lambda/ 6 M^3}$.
 The effective Planck scale is then given by 
 \be
 M_{P\ell}^2 = {M^3\over k}\left(1- e^{-2kr\pi}\right). 
 \ee

The effect of this metric on the brane fields parameters 
is non trivial. Lets consider for instance the Higgs action for the
brane at the end of the space, where we assume all SM fields are fixed, it
is given by
 \be
 S_H =
\int\! d^4x \sqrt{-g}\ e^{-4kr\pi} \left [e^{2kr\pi} g^{\mu\nu}
     \partial_\mu H\partial_\nu H - \lambda\left(H^2 - \hat
     v_0^2\right)^2\right].
 \ee  
After introducing the normalization $H\rightarrow e^{kr\pi} H$ that
recovers the canonical kinetic term, the above action becomes
 \be
 S_H = \int\! d^4x \sqrt{-g}  \left[ g^{\mu\nu}
     \partial_\mu H\partial_\nu H - \lambda\left(H^2 - 
     v^2\right)^2\right],
 \ee
where the vacuum $v= e^{-kr\pi}\hat v_0$. Therefore, by choosing 
$kr\sim 12$, the physical mass of the Higgs field, its vacuum, and, thus
all the SM masses appear at the TeV scale without needness or a
large hierarchy on the radius, even though the original mass parameter
$\hat v_0\sim M\sim M_{P\ell}$. Notice that on the contrary, any field
located on the other brane will get a mass of the order of $M$. Moreover, 
it also implies that no new particles exist in our world with
masses larger than TeV.

\subsection{Some about the Phenomenology}

Despite this impressive property of the model, which  has attracted a lot
of attention.  It also  turns out that the effective cut-off of the
theory is also at the TeV scale. Indeed, all high order operators get
now suppressed just by  $m_0\sim e^{-kr\pi} M$, that is at the TeV
range. The reason is as follows.  Any operator of dimension $n$,
$\Theta_n$, is originally suppressed by the fundamental scale $M\sim
M_{P\ell}$. However, under the change on the normalization of the fields,
$\Theta_n\rightarrow e^{-nkr\pi}\Theta_n$. Therefore, on the  effective
theory we get the large enhancement $M\rightarrow m_0$. So far no definite
solution to the dangerous presense of those operators exist, tough one
might impose global symmetries again. On the other hand, this scenario
has not yet been realized on the context of string theory. 

Kaluza Klein modes in this model also get masses at the TeV scale, and
the zero modes remain massless. They may get mass from brane contributions. 
The radion, however, 
will pick up a mass from the stabilization
potential~\cite{gw,csaki}. The intriguing possibility that the extra 
space could be actually infinite was presented in~\cite{rsth}.
Some phenomenological bounds on the effect of
KK gravitons are found in~\cite{rizzo}. Neutrino masses were analyzed
in~\cite{grossman}. Here, only the zero modes  get tiny masses.  Since
the KK modes are heavy,
they do not participate on the mixings. Thus, there is
not a clear 
experimental signature for this models in the neutrino sector. A further
application of this model, and its extensions, could be a possible
resolution of the cosmological constant problem~\cite{cc}

Cosmology on this models has received a lot of attention. It
was early observed~\cite{binetruy} that the Hubble parameter
$H$ is proportional to the density on the brane, $\rho$,
instead of the usual $H\sim\sqrt\rho$ of the standard big bang cosmology.
This could be disastrous for late cosmology and BBN. Nevertheless, at low
energies the leading order on $H$ has the right
behaviour~\cite{csaki,cline}.  A big deal of work has been devoted to
further study those ideas~\cite{cosmo,cosmo2}. 
Density and tensor perturbations in 
brane-world cosmology was considered
in  Ref.~\cite{perturb}.

\subsection{Radion stabilization}

Other interesting aspect of the RS framework is the recent considerations
of the stabilization mechanism provided by a bulk scalar field. 
The idea is as
follows. Consider the bulk scalar action
 \be 
 S_\phi = {1\over 2}\int\! d^5x \sqrt{-G}
 \left(G^{AB}\partial_A\phi\partial_B\phi - m^2\phi^2\right).
 \ee
Let us assume that the scalar field satisfies certain boundary conditions
associated to its couplings to the visible and hidden branes (at $y=0$ and
$y=\pi$ respectively). For instance
 \be
 S_{h,v} = -\int\!d^4x\sqrt{-g_{h,v}}\ \lambda_{h,v} \left(\phi^2 -
 v_{h,v}^2\right)^2  . 
 \ee
Those terms cause $\phi$ to develop a $y$-dependent vacuum which is
determined classically by solving the equation of motion. Inserting this
solution into $S_\phi$ and integrating over $y$ yields an effective
potential for the radius  of the form~\cite{gw}
 \be 
 V_{eff}(r) = 4 k e^{-4kr\pi}\left( v_v - v_he^{-\epsilon
 kr\pi}\right)^2 ;
 \ee
where $\epsilon= m^2/4k^2\ll 1$. This has a minimun at 
 \be 
 kr = \left({4\over \pi}\right) {k^2\over m^2}\ln\left[{v_h\over
 v_v}\right] \ . 
 \ee
With the logarithmic part of the order of one, we just need $k^2/m^2$ of
order 10 to get the right order in $kr$. Notice that despite the
fact that we are only  passing  the small hierarchy on $kr$ 
to the ratio among $k$ and $m$, this model provides a stabilization
potential that also gives a mass for the radion in the TeV range or
so~\cite{gw,csaki}.

\section{Concluding remarks}

The  wave induced by the seminal works in~\cite{dvali,rs} has generated a
big industry that is still growing. Several aspects of this higher
dimensional models have been investigated in the recent years and  more
is yet to come.   As any new field, the study of physics on large extra
dimensions is still confronting several criticism that eventually have
possed serious challenges. So far the field have survived to many of
these test, though several open questions remain. Several ingenious
applications have attracted the attention of the community and created new
directions of research. We mentioned along these notes what we believe
are some of the more interesting applications of the idea. However we must stress that
some of them are still disconnected of the other parts. It is fair to say
that yet a definite and comprehensive  model  of our world in this
framework  does not exist. So far, only  separate pieces of the puzzle have
been analyzed. There are many unsolved problems not  just of technical
nature but of fundamental nature. There is the problem of stabilization
of the large 
extra dimensions; connecting all the phenomenology (neutrino masses, proton
decay, etc.) in a single picture; and studying
the impact of this scenario in other  well established areas of
physics.  There is for instance the realization of  Randall Sundrum
and other models from string theory. 

Besides the hope of observing deviations in the Newton law at small
distances in the near future,  neutrino physics on this models seems very
promising too. On the other hand, it is possible that the collider
experiments may only increase the lower bounds on the fundamental
scale (as for other models of new physics), although we could
also discover the first signals of extra dimensions on the next colliders. 
Besides, let us mention that
while in these theories  it seems possible to maintain gauge coupling
unification, it is not yet clear whether it leads to a compelling scenario
that may provide any light on the origin of other SM parameters.
Certainly it seems clear  that several other mechanisms must be invoked
in contrast with the progress  that it has been made
on  four dimensional theories~\cite{pati}.  
There is, however,  the hope
that the years to come helps us either on conforming a more accurate
picture or, perhaps on   ruling  out these ideas.
Surely, the coming years will see a lot more on this topic,
and perhaps new ideas and results could set it on the side of the well
established world of physical theories. Meanwhile, most of the present
results remain speculative, although well motivated.  As always, Nature
has the last word.

The present notes have been prepared intending to be a first introductive
guide to the newborn field of models in large (and short) extra
dimensions.  Unable to make reference to all existing works in the area,
we have tried to collect those  we believed are relevant for our goal,
although  some important works
could have been omitted.  We advise the interested reader to consult the
more extended literature that exist already.
 
\null

{\bf Acknowledgments}.
I would like 
to thank the warm hospitality of the  Particle Physics Group at the
UMCP during the last two years. I have been beneficed by the 
collaboration with R.N. Mohapatra, S. Nandi, C.A. de S. Pires,  and L.
D\'{\i}az,  and by the discussions on this topics with J.C. Pati,  
S. Nussinov, M. Luty, E. Pont\'on,  C. Csaki, Z. Chacko,  K. Dienes, H.H.
Garc\'{\i}a-Compe\'an and G. Wolf,
to whom I am very thankful. This work was supported in part by CONACyT
(M\'exico). 




 \begin{table}[h]
  \begin{tabular}{||c|c|c|c||}
 Process & Background & M limit & Collider \\	   \hline
 $e^+e^-\rightarrow \gamma G$ &  $e^+e^-\rightarrow\gamma\bar\nu\nu$
       & 1 TeV & L3 \\
 $e^+e^-\rightarrow ZG $ & $ e^+e^-\rightarrow Z\bar\nu\nu$ &  
       $\bigg\{$\begin{tabular}{c} 515 GeV \\ 600 GeV \end{tabular} & 
       \begin{tabular}{c} LEPII \\ L3 \end{tabular} \\
 $Z\rightarrow \bar ffG$ &
 $Z\rightarrow  \bar ff\bar\nu\nu$ & 0.4 TeV & LEP 
 \end{tabular}
 \caption{Collider limits for the fundamental scale $M$. Graviton
 Production.}
 \end{table}

 \begin{table}[h]
  \begin{tabular}{|| c| c| c ||}
 Process  & M limit & Collider \\
       \hline
 $e^+e^-\rightarrow f\bar f$ & 0.94 TeV &  Tevatron \& HERA\\
 $e^+e^-\rightarrow \gamma \gamma,WW, ZZ$ & 
       $\bigg\{$\begin{tabular}{c}0.7 -- 1 TeV \\ 0.8 TeV\end{tabular}&
	      \begin{tabular}{c} LEP\\ L3 \end{tabular} \\
 All above & 1 TeV & L3\\
  Bhabha scattering & 1.4 TeV & LEP \\
 \begin{tabular}{c} $q\bar q\rightarrow \gamma \gamma $ \\
  $gg\rightarrow \gamma \gamma $ \end{tabular}$\bigg\}$  & 0.9 TeV & CDF \\
 \hline
 \end{tabular}
 \caption{Collider limits for the fundamental scale $M$. Virtual Graviton
 exchange}
 \end{table}

\end{document}